\documentclass[12pt]{report}

\usepackage[utf8]{inputenc}
\usepackage{graphicx}
\usepackage{amsmath, amssymb}
\usepackage[left=3cm, right=3cm, top=3cm, bottom=3cm]{geometry}
\usepackage{titlesec}
\usepackage{listings}
\usepackage[dvipsnames]{xcolor}
\usepackage[utf8]{inputenc}
\usepackage{mathtools}
\usepackage{tikz}
\usetikzlibrary{shapes,snakes,arrows.meta}

\newcommand{\KP}[1]{
  \begin{tikzpicture}[baseline=-\dimexpr\fontdimen22\textfont2\relax]
  #1
  \end{tikzpicture}
}
\newcommand{\KPA}{
  \KP{\filldraw[color=red, fill=none, thick] circle (0.3);}%
}
\newcommand{\KPB}{
  \KP{
    \draw[color=red,thick] (-0.3,0.3) -- (0.3,-0.3);
    \draw[color=red,thick] (-0.3,-0.3) -- (-0.05,-0.05);
    \draw[color=red,thick] (0.05,0.05) -- (0.3,0.3);
  }
}

\newcommand{\KPC}{
  \KP{
    \draw[color=red,thick] (-0.3,0.3) .. controls (0,-0.05) .. (0.3,0.3);
    \draw[color=red,thick] (-0.3,-0.3) .. controls (0,0.05) .. (0.3,-0.3);
  }
}
\newcommand{\KPD}{
  \KP{
    \draw[color=red,thick] (-0.3,-0.3) .. controls (0.05,0) .. (-0.3,0.3);
    \draw[color=red,thick] (0.3,-0.3) .. controls (-0.05,0) .. (0.3,0.3);
  }
}

\titleformat{\chapter}[display]
  {\normalfont\huge\bfseries}{\chaptertitlename\ \thechapter}{20pt}{\Huge}
\titleformat{\chapter}[hang] 
  {\normalfont\huge\bfseries}{}{0pt}{\Huge}
\titleformat{\section}
  {\normalfont\Large\bfseries}{}{0pt}{}
\titleformat{\subsection}
  {\normalfont\medium\bfseries}{}{0pt}{}

\begin{document}

\begin{titlepage}
    \centering
    \vspace*{1cm}
    \Large\textbf{Development of a Python-Based Software for Calculating the Jones Polynomial: Insights into the Behavior of Polymers and Biopolymers}\\
    \vspace{0.3cm}
    \normalsize by\\
    \vspace{0.3cm}
    \Large Caleb Musfeldt\\
    \vspace{0.5cm}
    \normalsize A thesis presented for the Honors Degree of Bachelor's of Science of Data Science\\
    \vspace{0.5cm}
    \normalsize Advisor: Eleni Panagiotou, Ph.D.\\
    \vspace{0.25cm}
    \normalsize Committee Members: Eleni Panagiotou, Ph.D. and Andrea Richa, Ph.D.\\
    \vfill
    \begin{figure}[h]
    \centering
    \includegraphics[width=0.5\textwidth]{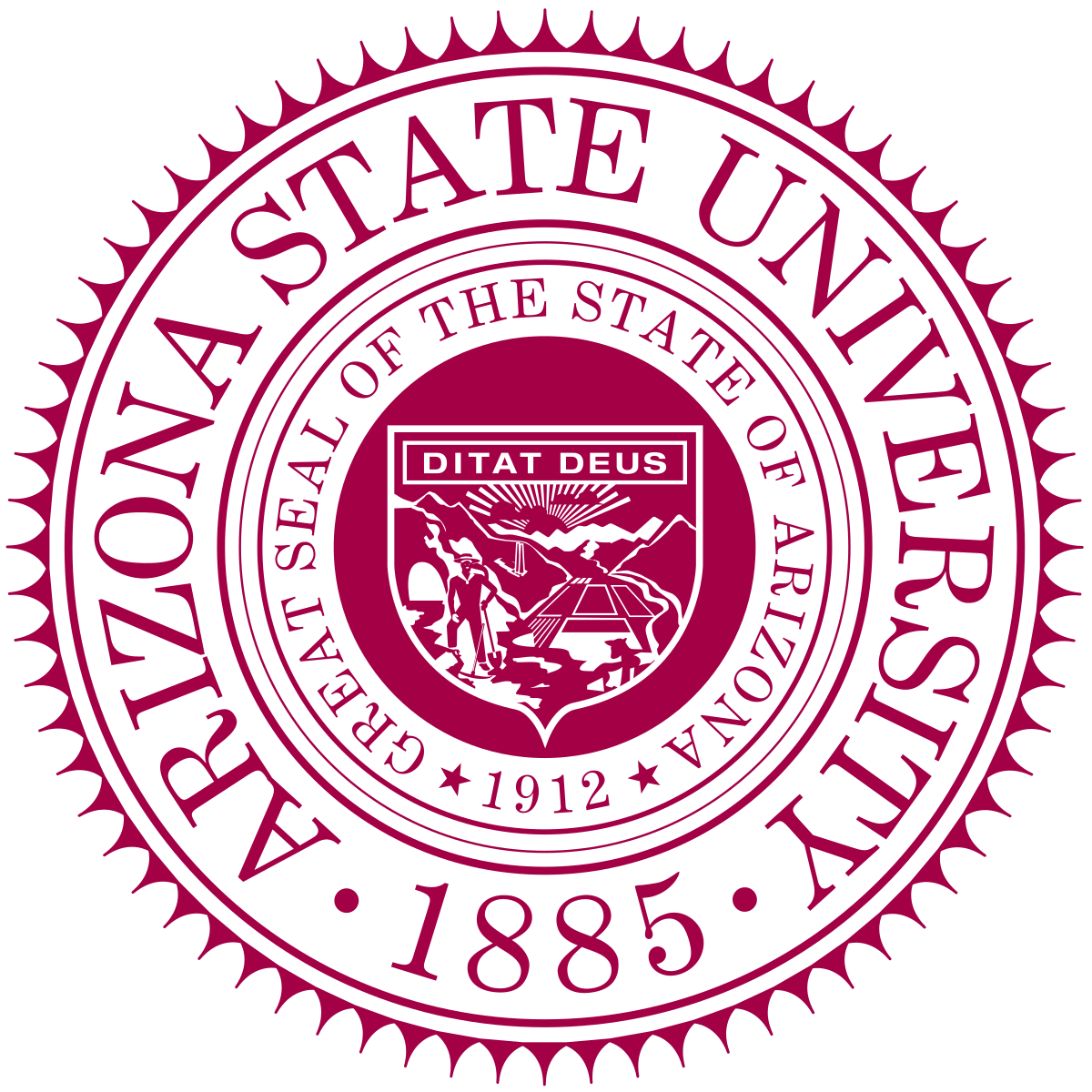}
    \end{figure}
    \vfill
    \normalsize Submitted: April 12, 2024\\
    \vspace{0.8cm}
    \normalsize School of Mathematics and Statistical Sciences\\
    \normalsize Barrett, the Honors College\\
    \normalsize Arizona State University\\
\end{titlepage}

\begin{abstract}
This thesis details a Python-based software designed to calculate the Jones polynomial, a vital mathematical tool from Knot Theory used for characterizing the topological and geometrical complexity of curves in \( \mathbb{R}^3 \), which is essential in understanding physical systems of filaments, including the behavior of polymers and biopolymers. The Jones polynomial serves as a topological invariant capable of distinguishing between different knot structures. This capability is fundamental to characterizing the architecture of molecular chains, such as proteins and DNA. Traditional computational methods for deriving the Jones polynomial have been limited by closure-schemes and high execution costs, which can be impractical for complex structures like those that appear in real life. This software implements methods that significantly reduce calculation times, allowing for more efficient and practical applications in the study of biological polymers. It utilizes a divide-and-conquer approach combined with parallel computing and applies recursive Reidemeister moves to optimize the computation, transitioning from an exponential to a near-linear runtime for specific configurations. This thesis provides an overview of the software's functions, detailed performance evaluations using protein structures as test cases, and a discussion of the implications for future research and potential algorithmic improvements.
\end{abstract}

\tableofcontents

\chapter{Introduction}
Topology is a fundamental branch of mathematics that studies the properties of space that are preserved under continuous deformations, such as stretching, crumpling, and bending, but not tearing or gluing. It explores the qualitative aspects of geometry rather than focusing only on quantitative measurements.
\\ \\
Knot theory, a branch of topology, focuses in the classification of curves in space. The field has found itself in the heart of biological sciences, as many physical systems are composed by filamentous structures that can be modeled by curves in space. Through studying the structure of biological phenomena, science gains insights into the fundamental principles of life processes. However, there are many difficulties in applying knot theory to physical systems. For example, measuring entanglement of open curves in \( \mathbb{R}^3 \) (as most physical filaments are) has become possible only recently, and the computational time of most of topological functions is prohibiting their application to real physical systems. 
\\ \\
In this thesis, we focus on an algorithm for the computation of the Jones polynomial of both open and closed chains and methods for improving its computational time. The Jones polynomial is a very useful tool that enables to distinguish different knot and link types very well, even though it is not a complete invariant. Even though its original definition requires a background in topology, it also has a practical combinatorial definition in terms of knot and link diagrams, which makes it easy to understand. However, its combinatorial definition requires an exponential time computation that depends on the number of crossings in a diagram. It is worth mentioning that systems of physical filaments can easily have more than 100 crossings in a diagram.
\\ \\
The thesis is organized as follows: Chapter 2 discusses basic concepts in knot theory, Chapter 3 presents the definition of the Jones polynomial, Chapter 4 gives an overview and comparison of existing software for computing the Jones polynomial. Chapter 5 introduces the software we use and develop in this thesis, Chapter 6 discusses specific aspects of the software that enable a decrease in its computational time. Chapter 7 discusses the performance of the new algorithm and Chapter 8 presents the conclusions and future work directions that stem from our analysis. 

\chapter{Knot Theory}
\begin{figure}[h]
\centering
\includegraphics[width=0.5\textwidth]{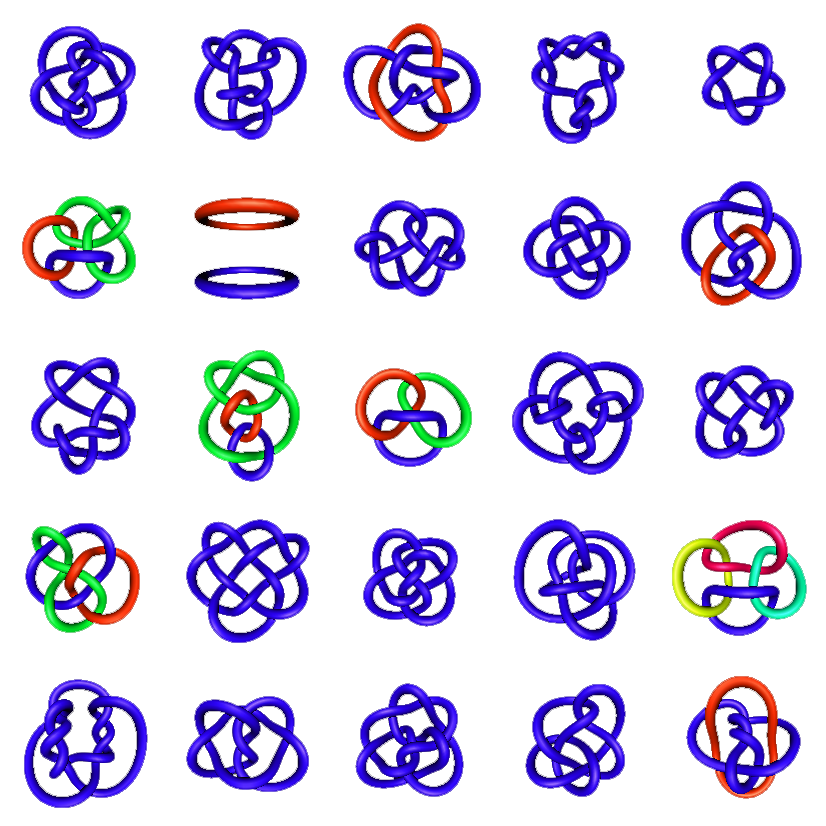}
\caption{Examples of mathematical knots and links. The knots and links we focus on live in 3 dimensions. It is easier, however, to use their 2 dimensional diagrams with over/under information at the crossing to study the 3D knots and links.   Image from \cite{BestvinaND}}
\label{fig:knots}
\end{figure}

A knot is essentially a simple closed curve in \( \mathbb{R}^3 \). Think of a tangled piece of string with its ends joined together so that it cannot be untangled without cutting the string, which can be illustrated in \ref{fig:knots}. The simplest form of a knot is known as the "unknot" or "trivial knot," which can be deformed to a simple loop that does not have any twists or tangles \cite{Adams2004}. 
\\ \\
A key point about knots is that they consist of a single, continuous loop. A link, on the other hand, consists of two or more knots (or loops or components) that are entangled in some way but do not intersect. Each component of the link is a knot, and when there are multiple knots intertwined they become a link. A simple example of a link is two circles linked together like a chain, known in Knot Theory as a "Hopf link". 
\\
\\
Knot Theory classifies these structures through the concept of topological equivalence. That is, knots (or links) are said to be equivalent when they can be deformed to each other without allowing cutting and pasting. Topological invariants, are quantities or features of the knots (or links) that remain unchanged under such deformations. This theory has provided a mathematical framework for understanding tangled structures while also finding applications in analyzing the entanglement of the many architectures of physical filament.

\section{Reidemeister Moves}

\begin{figure}[h]
\centering
\includegraphics[width=0.5\textwidth]{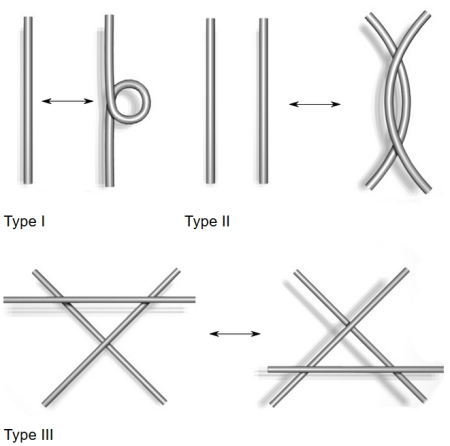}
\caption{Reidemeister moves are local moves on diagrams. By applying them multiple times we can go from any knot diagram to any other diagram of the same knot. Image from \cite{Baez2018}}
\label{fig:reidemeister}
\end{figure}

Since they are easier to work with, we often study knots and links by using diagrams. These are 2D projections of knots that preserve the information of over and under strand at each crossing. The notion of topological equivalence of knots in \( \mathbb{R}^3 \) translates to equivalence of knot diagrams. Invariants are then quantities computed on a diagram that do not change for diagrams that represent the same knot. This becomes easier to define through Reidemeister moves.
\\ \\
Reidemeister moves, introduced by Kurt Reidemeister in 1927 \cite{Reidemeister1927}, are operations on a knot (or link) diagram that do not change the knot's topological properties and these 3 moves are sufficient to show equivalence of any two diagrams of the same knot. The three moves are visualized in \ref{fig:reidemeister}.
\\ 

\chapter{Jones Polynomial}
The Jones Polynomial is an invariant introduced by Vaughan Jones in 1984 \cite{Jones1985}. It assigns a polynomial to each oriented knot or link. The development of the Jones Polynomial was motivated by the need for a more robust method of classifying knots beyond previous invariants.
\\ \\
We will define the Jones polynomial as the normalized bracket polynomial. For this, we will introduce the bracket polynomial and the writhe of a diagram.

\section{Bracket Polynomial}
The bracket polynomial, also known as the Kauffman bracket polynomial for Louis Kauffman's introduction of it in 1987 \cite{Kauffman1987}, is used to derive the Jones polynomial. We compute this on a knot diagram by ``smoothing'' its crossings. The bracket polynomial is not a topological invariant. It becomes one (the Jones polynomial) after normalization by the Writhe, which we will discuss next. For a knot diagram $D$, the bracket polynomial is defined by:

\begin{enumerate}
\item{Trivial Link:}
  $\left\langle\KPA\right\rangle=1$
\item{Disjoint Union:}
  $\left\langle D \cup \KPA\right\rangle=(-A^{2}-A^{-2})\langle D\rangle$
\item{Skein Relation:}
  $\left\langle\KPB\right\rangle=
  A\left\langle\KPC\right\rangle + A^{-1} \left\langle \KPD \right\rangle$
\end{enumerate}

\section{Writhe}
The writhe, commonly portrayed as either $w(D)$ or $Wr$, is a number that quantifies the amount of times a knot turns around itself. It is defined by first assigning a sign to each crossing of an oriented knot diagram. The writhe is defined as the algebraic sum of the signs of the crossings in the diagram.
\\
\begin{center}
\begin{tikzpicture}[scale=1.5, every node/.style={scale=1}]
    \begin{scope}
        \draw (0.25,0.25) circle [radius=0.4cm];
        \draw[->,red] (0.5,0) -- (0,0.5);
        \draw[line width=1mm,white] (0,0) -- (0.5,0.5);
        \draw[->,red] (0,0) -- (0.5,0.5);
        \node at (0.25,-0.4) {\small $w(D) = +1$};
    \end{scope}
    
    % Negative crossing
    \begin{scope}[xshift=2cm]
        \draw (0.25,0.25) circle [radius=0.4cm];
        \draw[->,red] (0,0) -- (0.5,0.5);
        \draw[line width=1mm,white] (0.5,0) -- (0,0.5);
        \draw[->,red] (0.5,0) -- (0,0.5);
        \node at (0.25,-0.4) {\small $w(D) = -1$};
    \end{scope}
\end{tikzpicture}
\end{center}

\section{The Normalized Bracket Polynomial}
The Jones polynomial $V_D(t)$ for a knot diagram, can be derived from the bracket polynomial $\langle D \rangle$ and the writhe $w(D)$ using the relationship:
\[
V_D(t) = \left((-A^3)^{-w(D)}\right) \langle D \rangle
\]
where the substitution $A=t^{-1/4}$ is made after the calculation for further simplification.
\\ \\
One can prove that the Jones polynomial is invariant under Reidemeister moves. Thus, it is invariant under topological equivalence.

\section{The Jones Polynomial of Open Curves in \( \mathbb{R}^3 \)}

Open curves in \( \mathbb{R}^3 \) can also tie open knots and links, but they cannot be classified under topological equivalence, as they all belong to the same class (they can be deformed to one another without cutting and pasting). Similarly, an open curve in \( \mathbb{R}^3 \) cannot be studied by one of its diagrams, as different projections can capture different aspects of both its topology and its geometry. This is why, stronger methods are needed. In \cite{Panagiotou2020b} a new method was introduced that allows the rigorous definition of the Jones polynomial of open curves in \( \mathbb{R}^3 \). 
\\ \\
Let $l$ denote a curve in \( \mathbb{R}^3 \). Let $(l)_{\vec{\xi}}$ denote the projection of $l$ on a plane with normal vector $\vec{\xi}$. 
\\ \\
The normalized bracket polynomial of $l$ is defined as: 

\begin{equation}\label{avnk}
f_{K(l)}=\frac{1}{4\pi}\int_{\vec{\xi}\in S^2}(-A^3)^{-wr((l)_{\vec{\xi}})}\langle (l)_{\vec{\xi}}\rangle dS
\end{equation}
\\
where the integral is over all vectors in $S^2$ except a set of measure zero (corresponding to irregular projections).
\\ \\
For open curves in \( \mathbb{R}^3 \) the Jones polynomial is a continuous function of the curve coordinates and tends to the classical topological invariant when the endpoints tend to coincide.

\chapter{Existing Methods and Software}
There are several computational packages that enable the computation of the Jones polynomial in different settings. Below we overview some of these.
\\ \\
\textbf{Mathematica}
\\ \\
The Mathematica Knot Theory Package \cite{KnotTheory2009} allows users to input knots in various formats (typically as 2D Gauss code) and calculate the Jones Polynomial via symbolic manipulations, such as the Kauffman Bracket approach. Outputs include the Jones Polynomial as a Laurent polynomial, suitable for further Mathematica-based analysis. This package only works with closed knots and links.
\\ \\
\textbf{Sage Math}
\\ \\
Sage Math \cite{BuzunarizJamadagni2024}, an open-source software, supports Knot Theory and the calculation of the Jones polynomial. It allows for knot representations and uses several algorithms for polynomial computation. The resulting Jones Polynomial, is also presented as a Laurent polynomial and only works with closed knots and links.
\\ \\
\textbf{Topoly}
\\ \\
Topoly \cite{Dabrowski-Tumanski2020} focuses on the topological analysis of biological polymers. It accepts data in formats like PDB files and employs algorithms suited to biological structures for calculating the Jones Polynomial. Topoly works with closed curves or uses a closure approximation method for open curves.
\\ \\
\textbf{TEPPP (Topological Entanglement in Proteins, Polymers and Periodic Systems) }
\\ \\
TEPPP \cite{Panagiotou2022} is a software written in C++ that is designed to compute the Jones polynomial for both open and closed curves, demonstrating versatility in handling different types of topological configurations. The original version of TEPPP was developed to work exclusively with single components, meaning it could process only one knot or open curve in \( \mathbb{R}^3 \) at a time. This focus allows for precise calculations specific to individual topological entities, which is essential in studies where the properties of isolated curves are under examination, or when average single chain properties are analyzed. Significantly, TEPPP offers a methodological advantage by being the only available software that eliminates the need for closure schemes or 2-dimensional approximation schemes, by employing the novel rigorous definition of the Jones polynomial of open curves in \( \mathbb{R}^3 \) \cite{Panagiotou2020b}.

\chapter{Software Overview}
The software discussed in the following come from original code by Kasturi Barkataki and Eleni Panagiotou \cite{BarkatakiPanagiotou2024}, adapted by code from  Herschberg, T., Pifer, K. and Panagiotou, E. \cite{Panagiotou2022} (TEPPP).
\\ \\
The code consists of the following main functions:

\section{$jones\_knot.py$}
The $jones\_knot.py$ script is designed to calculate the Jones polynomial for a given knot or linkoid represented in \( \mathbb{R}^3 \). This Python uses the functionality provided by the $JONES.py$ module, specifically importing the $expected\_jones()$ function, which is essential for carrying out the polynomial calculations.
\\ \\
$jones\_knot.py$ uses a variable $n$ that defines the number of projections used for the computation of the Jones polynomial of open curves. For a knot, this should be set to 1. One can set the variable n to 1 for open curves as well if needed to check particular examples.
\\ \\
The script defines $input\_knot$ as a numpy array containing the coordinates that delineate the structure of the knot or open curve in \( \mathbb{R}^3 \). These coordinates are then passed as an argument to the $expected\_jones()$ function along with $n$.
\\ \\
After computing the Jones polynomial, $jones\_knot.py$ prints out the result. Additionally, the script calculates and displays the computation time, offering insight into the performance of the polynomial calculation process as well helping to manage larger simulations.

\section{$JONES.py$}
The script begins by setting the $closed$ variable, which dictates whether the input curves are open or closed. The main function, $expected\_jones()$, initializes dictionaries to keep track of directed indices and coordinates of the points making up the curves.
\\ \\
For generating projection vectors necessary for the polynomial calculations, $expected\_jones()$ uses the $fibonacci\_sphere()$ function to create $n$ vectors uniformly distributed on the unit sphere if $n$ is greater than one. If only a single projection is needed, a specific projection direction is chosen or a random one may be generated. With these vectors, along with the pre-defined coordinates and indices, the script proceeds to the $get\_jones\_poly()$ function where the actual computation of the Jones polynomial begins.
\\ \\
Within this function, the $crossing\_matrices\_bool()$ is utilized to generate matrices that describe the crossings in the knot diagram accurately. These matrices are then used in the $simplification()$ function, which applies Reidemeister moves to simplify the knot diagram, an essential step in reducing computational costs. Following $simplification()$, the $get\_partial\_poly()$ function computes the bracket polynomial and $get\_writhe()$ calculates the writhe.
\\ \\
The results from all projected vectors are combined, and the script normalizes these results based on the number of successful calculations to account for any projections that may have been discarded due to irregularities or complexities in the knot diagram. Finally, $JONES.py$ returns the averaged Jones polynomial. 

\section{$functions.py$}
Contains helper functions that are called throughout JONES.py. Notable functions include the Reidemeister moves and all of the Jones polynomial operators. This script also includes the simplification function where we define the order of our Reidemeister moves.

\chapter{Implementations of More Efficient Methods}
In order to make the software as efficient as possible we developed a computational method that "splits" the knot into two components and parallelizes the computations of the bracket polynomial for each before "gluing" the components back together, a method developed by Barkataki and Panagiotou \cite{BarkatakiPanagiotou2024}. 
\\ \\
We also needed simplification methods that are efficient and reliable. We developed functions for each Reidemeister move to simplify more complex knots to ensure the Jones polynomial is obtained as quickly as possible. The order for Reidemeister moves during testing was RM1, RM2, RM3, RM1, RM2.

\section{Splitting and Gluing}
\textbf{Splitting}
\\ \\
First, we determine if there are any crossings in the knot/link by checking if the crossing matrix ($bool\_mask$) has any $True$ values. If no crossings are present, the computation does not proceed further. The knot/link is then split into two linkoids, $L1$ and $L2$, based on arcs determined to define $L1$. This involves identifying arcs for $L1$ and generating $inds\_L1$ (index mappings for $L$1) and similarly for $L2$, sorting arcs for $L2$ to maintain order. We prepare crossing, over/under, and right/left matrices ($BM\_L1$, $OU\_L1$, $RL\_L1$, etc.) for both linkoids by setting crossings involving the other linkoids arcs to $False$. We make modifications to the indices of $L1$ and $L2$ are made to match the expected format for the $get\_partia\_poly()$ function. We initialize empty lists $states1$ and $states2$ for tracking states in the partial polynomial computations. The script then calculates partial bracket polynomials for $L1$ and $L2$ parallely using $get\_partial\_poly()$, $parallel\_partial\_poly()$, and $ThreadPoolExecutor$ from \cite{Python2024}.
\\ \\
\textbf{Gluing and Final Bracket Polynomial Calculation}
\\ \\
To "glue" we combine the bracket polynomials of $L1$ and $L2$ to compute the final bracket polynomial of the original knot/link. This involves generating all combinations of terms from $L1$ and $L2$ polynomials. For each combination, summing the A-factors and constructing a system to identify connected components (or loops). Using the $J\_add()$ and $J\_mult()$ functions, along with $dfactor$, to aggregate the contributions to the bracket polynomial. The final step adjusts the computed polynomial for the writhe of the knot/link. This is done using $J\_mult()$ and adjusting the result based on the writhe, where 3*(-Wr) adjusts the polynomial's exponent and (-1)**abs(Wr) its sign based on the writhe's absolute value.
\\ \\
Code block from $get\_jones\_poly()$, theoretical method developed in \cite{BarkatakiPanagiotou2024}.
\begin{lstlisting}
if np.count_nonzero(bool_mask)/2. >0.:
        ## SPLITTING OF THE KNOT INTO 2 LINKOIDS L1 and L2
        ## Arc defining L1
        Arc=find_arc(bool_mask,inds)
        L1_arcs=Arc[0]+Arc[1]
        inds_L1 = {k: v for k, v in inds.items() if k in L1_arcs}
        inds_L2 = {k: v for k, v in inds.items() if k not in L1_arcs}
        L2_arcs = list(inds_L2.keys())
        L2_arcs.sort()
        BM_L1= np.copy(bool_mask)
        OU_L1= np.copy(over_or_under)
        RL_L1= np.copy(right_or_left)
        for i in L2_arcs:
            BM_L1[i,:]=False
            BM_L1[:,i]=False
            OU_L1[i,:]=False
            OU_L1[:,i]=False
            RL_L1[i,:]=False
            RL_L1[:,i]=False
        BM_L2= np.copy(bool_mask)
        OU_L2= np.copy(over_or_under)
        RL_L2= np.copy(right_or_left)
        for i in L1_arcs:
            BM_L2[i,:]=False
            BM_L2[:,i]=False
            OU_L2[i,:]=False
            OU_L2[:,i]=False
            RL_L2[i,:]=False
            RL_L2[:,i]=False
        ## The bracket polynomials of the L1 and L2 linkoids.
        ## First bring inds_L1 and inds_L2 to desired format.
        modf_inds_L1 = {}
        modf_inds_L2 = {}
        for i in inds_L1:
            modf_inds_L1[i]=[inds_L1[i]]
        for i in inds_L2:
            modf_inds_L2[i]=[inds_L2[i]]
        states1=[]
        states2=[]

        with ThreadPoolExecutor(max_workers=2) as executor:
            # Submit tasks for parallel execution
            future_L1 = executor.submit(parallel_partial_poly, BM_L1, OU_L1, RL_L1, inds_L1, modf_inds_L1, states1)
            future_L2 = executor.submit(parallel_partial_poly, BM_L2, OU_L2, RL_L2, inds_L2, modf_inds_L2, states2)

            L1_pp = future_L1.result()
            L2_pp = future_L2.result()

        if L1_pp!=None:
            print("L1 done")
        print('======================')
        if L2_pp!=None:
            print("L2 done")

        ## GLUING back information from L1_pp and L2_pp. Get bracket polynomial as state sum terms.
        Bkt_poly={0:0}
        for L in L1_pp:
            for M in L2_pp:
                Af=L[0]+M[0]
                S={}
                for i in range(len(L[1])):
                    S[i]=L[1][i]
                for j in range(len(M[1])):
                    S[j+len(L[1])]=M[1][j]
                K={}
                for i in range(len(L[1])+len(M[1])):
                    for j in range(len(L[1])+len(M[1])):
                        if S[i].intersection(S[j])!= set([]):
                            try:
                                K[i].append(j)
                            except:
                                K[i]=[j]
                Bkt_poly=J_add(Bkt_poly,J_mult({Af:1},dfactor(len(Loops(K)[1])-1)))
        return J_mult(Bkt_poly,{3*(-Wr):(-1)**abs(Wr)})
\end{lstlisting}

\section{Reidemeister Moves}
\textbf{$RM1()$}
\\ \\
The process begins with $bool\_mask$ and $inds$ as the inputs. The procedure involves iterating over each arc, denoted as $a$, within the $inds$, and identifying the subsequent arc it connects to, labeled as $b$. During each iteration, the script checks if a and b have crossings and whether the pair [$b$, $a$] is marked as $True$ in $bool\_mask$. If both conditions are met, Reidemeister Move 1 is applied by setting both $bool\_mask[a, b]$ and $bool\_mask[b, a]$ to $False$, effectively removing these crossings marking both $a$ and $b$ as processed. The iteration then moves to the next arc. If the current arc does not meet the move criteria or lacks crossings, the loop continues without modifications, and exits. The output of this process is an updated $bool\_mask$.
\begin{lstlisting}
    def RM1(bool_mask, inds):
        # Initialize an empty list to keep track of processed indices.
        Bin = []
        
        # Iterate over each arc in the provided list of indices.
        for a in inds:
            # Check if the current arc has not been processed yet.
            if a not in Bin:
                # Check if the arc has any crossing.
                if np.any(bool_mask[a, :]):
                    # Initialize a flag to stop the while loop.
                    flag = 0
                    # Get the next point the current arc travels to.
                    b = inds[a]
                    # Enter while loop
                    while flag == 0:
                        # Check if the arc that our first arc connects to has any crossings.
                        if np.any(bool_mask[b, :]):
                            # Check if the value at position [b, a] in the boolean mask is True, only necessary property for RM1.
                            if bool_mask[b, a] == True:
                                # Set BM to False to represent RM1
                                bool_mask[a, b] = False
                                bool_mask[b, a] = False
                                # Add both indices to the Bin list as processed.
                                Bin.append(a)
                                Bin.append(b)
                                # Set flag to 1 to exit the loop.
                                flag = 1
                            else:
                                # The arc we are checking does not meet the criteria to perform RM1, move on.
                                Bin.append(a)    
                                flag = 1
                        else:
                            # If no True value is found in the row corresponding to index b, try to update b to the next index in inds.
                            try:
                                b = inds[b]
                            except:
                                # If updating b fails due to an IndexError, exit the loop.
                                flag = 1
    
        return bool_mask
\end{lstlisting}
\textbf{$RM2()$}
\\ \\
The function starts with the $bool\_mask$, $over\_or\_under$, and $inds$. It will iterate over each arc $a$ within $inds$ to identify the first crossing, denoted as $c$, and the corresponding arcs $b$ and $d$ involved in this crossing. The simplification process checks if the pairs $a$ and $c$, as well as $b$ and $d$, can be simplified by Reidemeister Move 2. This is determined by verifying if $bool\_mask[b, d]$ is $True$ and whether their $over\_or\_under$ values match. If the criteria for the move are met, the $bool\_mask$ values between these arcs are set to $False$, representing the crossing being removed. The function then continues to explore remaining arcs, labeled as $e$, for potential simplification opportunities with arcs $a$ and $c$. This iterative checking and updating continue throughout the set of arcs until no further simplifications are found. The output is an updated $bool\_mask$.

\begin{lstlisting}
        def RM2(bool_mask, over_or_under, inds):
    # Initialize an empty list to track indices that have been processed.
    Bin = []
    
    # Iterate over each arc in inds.
    for a in inds:
        # Ensure the current index hasn't already been processed.
        if a not in Bin:
            # Retrieve the corresponding index from `inds` for the current index `a`.
            b = inds[a]
            # Check if there's any True value in the row of `bool_mask` corresponding to `a`.
            if np.any(bool_mask[a, :]):
                # Find the column index of the first True value in the row for `a`.
                c = np.argmax(bool_mask[a, :])
                # Retrieve the corresponding index from `inds` for the index `c`.
                d = inds[c]
                try:
                    # Check if the boolean mask at [b, d] is True and if the "over_or_under" values for [a, c] and [b, d] match.
                    if bool_mask[b, d] == True:
                        if over_or_under[a, c] == over_or_under[b, d]:
                            # If conditions are met, set the corresponding positions in `bool_mask` to False.
                            bool_mask[a, c] = False
                            bool_mask[c, a] = False
                            bool_mask[b, d] = False
                            bool_mask[d, b] = False
                            # Add the indices a, b, c, and d to the list of processed indices.
                            Bin.append(a)
                            Bin.append(b)
                            Bin.append(c)
                            Bin.append(d)
                except:
                    # Ignore errors, likely from indexing issues.
                    pass
                
                # If index `c` has not been processed, iterate over `inds` to find additional matches.
                if c not in Bin:
                    for e in inds:
                        # Ensure `e` hasn't been processed and is not the same as `a`.
                        if e not in Bin and e != a:
                            # Retrieve the corresponding index for `e`.
                            f = inds[e]
                            # Check if there's any True value in the row of `bool_mask` corresponding to `e`.
                            if np.any(bool_mask[e, :]):
                                # Find the column index of the first True value in the row for `e`.
                                g = np.argmax(bool_mask[e, :])
                                # Retrieve the corresponding index from `inds` for the index `g`.
                                h = inds[g]
                                # Check if `g` hasn't been processed and specific conditions are met.
                                if g not in Bin and (bool_mask[g, e] and over_or_under[a, c] == over_or_under[g, e]):
                                    # If conditions are met, set the corresponding positions in `bool_mask` to False.
                                    bool_mask[a, c] = False
                                    bool_mask[c, a] = False
                                    bool_mask[e, g] = False
                                    bool_mask[g, e] = False
                                    # Add the indices a, c, e, and g to the list of processed indices.
                                    Bin.extend([a, c, e, g])
    # Return the modified boolean mask after processing.
    return bool_mask
\end{lstlisting}
\textbf{$RM3()$}
\\ \\
The inputs include $bool\_mask$, $over\_or\_under$, and $inds$. First, the algorithm generates a list of crossing pairs $(i, j)$ where $over\_or\_under[i][j]$ is $True$, indicating that arc $i$ crosses over arc $j$. The function then creates combinations of three arcs to identify potential configurations for Reidemeister move 3. This involves analyzing sets of crossing triplets to check if the criteria for the move within these pairs are met. Once a valid move is identified, the $bool\_mask$ values for the original crossings with the strand are set to $False$. Subsequently, it updates the $bool\_mask$ for the new configurations to $True$, adding new crossings that result from the post-slide beneath or above our marked crossing. The output of this process is a $bool\_mask$ with updated crossing information that reflects a Reidemeister move 3 slide.
\begin{lstlisting}
        def RM3(bool_mask, over_or_under, inds):
    BM = bool_mask

    # Generate a list of every arcs that has a crossing.
    true_arcs = [(i, j) for i in range(len(over_or_under)) for j in range(len(over_or_under[i])) if over_or_under[i][j]]
    
    # Generate all possible combinations of 3 arcs to check for Reidemeister 3 conditions.
    comb_list = list(comb(true_arcs, 3))
    
    # Start iterating through the possible crossing combinations	
    for k in reversed(range(len(comb_list))):
        cross = list(comb_list[k])
    
        # Initialize lists to categorize arcs as 'under' or 'over'.
        under_list = []
        over_list = []
    
        # Create over and under lists to help evaluate the criteria for RM3.
        for i, j in cross:
            under_list.append(i)
            over_list.append(j)
    
        # Iterate through the arcs to find connected arcs in the over or under list.
        for i in range(len(over_list) - 1): # Need to duplicate so it will check over or under for sliding arc
            current_arc = over_list[i]
            # Iterate through each arc to see if any combination connects
            for j in range(i+1, len(over_list)):
                next_arc = over_list[j]
        
                # Check if two arcs are connected by inds.
                if inds.get(current_arc) == next_arc or inds.get(next_arc) == current_arc:
                    connected_arcs = [current_arc, next_arc]
            
                    # Define the crossing that our connected arcs will slide over or under.
                    slider_cross = [tup for tup in cross if not any(arc in tup for arc in connected_arcs)]
            
                    # Define the connected arcs that we will be sliding.
                    slider_arcs = [arc for arc in cross if arc[0] in connected_arcs or arc[1] in connected_arcs]
            
                    # Identify the indices for the arcs to be manipulated.
                    a, c = slider_arcs[i]
                    b = inds[a]
                    d = inds[c]
                    e, g = slider_arcs[i + 1]
                    f = inds[e]
                    h = inds[g]
                    i, k = slider_cross[-1]
                    j = inds[i]
                    l = inds[k]
            
                    # Update the BM to represent an RM3 move.
                    BM[a, c] = BM[c, a] = False
                    BM[e, g] = BM[g, e] = False
                    BM[c, k] = BM[k, c] = True
                    BM[g, i] = BM[i, g] = True

                    return BM
\end{lstlisting}

\chapter{Performance}
\section{Time Complexities}

In this section we discuss the computational complexity of the Jones polynomial code.
\\ \\
\textbf{Original Computational Complexity}
\\ \\
The complexity of the original method is exponential with respect to the number of crossings in the knot:
\[O(2^n)\]
where \(n\) is the number of crossings.
\\ \\
\textbf{Divide and Conquer with Parallel Computing}
\\ \\
By splitting the knot into smaller parts and applying parallel computing techniques, the complexity can be reduced:
\[O(2^{n/2})\]
This approach potentially reduces the exponent by half, assuming the knot is split into two equal parts. However, further study is required to obtain the exact order of computation time, as the pairing functions are also computationally expensive.
\\ \\
\textbf{Further Reduction with Reidemeister Moves}
\\ \\
Implementing Reidemeister moves to simplify the knot diagram further reduces the number of crossings:
\[O(2^{(n-k)/2})\]
where \(k\) (ranging from \(0\) to \(n\)) represents the number of crossings removed through these moves.
\\ \\
\section{Structures for Testing}

\begin{figure}
    \centering
    \includegraphics[width=1\textwidth, height=1\textwidth]{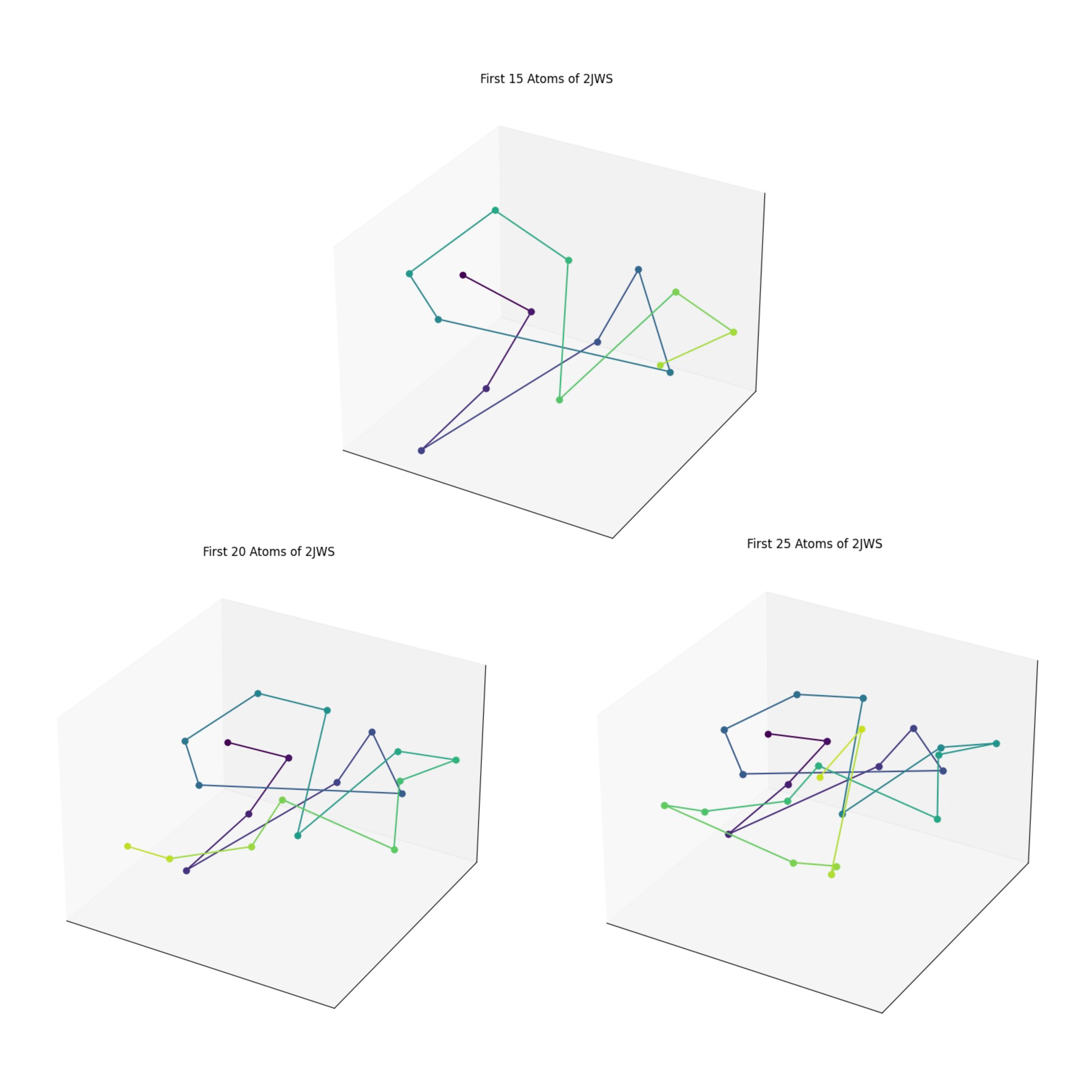}
    \caption{Structures used for testing the code in real systems. The structures were obtained from the protein 2jws.pdb.}
    \label{fig:chains}
\end{figure}

In addition to examples of specific knots, which we tested and behaved as expected, we also examined the efficiency and computational time of the code for proteins. The 2JWS protein structure was used for to test the execution times of the different methods. The protein coordinates were obtained from the Protein Data Bank \cite{Berman2000}. The tests were on the first 15 atoms, the first 20 atoms, and the first 25 atoms of the structure which can be seen in \ref{fig:chains}. These configurations had 6, 12, and 17 crossings from the projection [0, 0, 1] respectively.

\section{Execution Times}
\begin{figure}[h]
    \hspace{-75pt}
    \includegraphics[width=1.3\textwidth, height=0.8\textwidth]{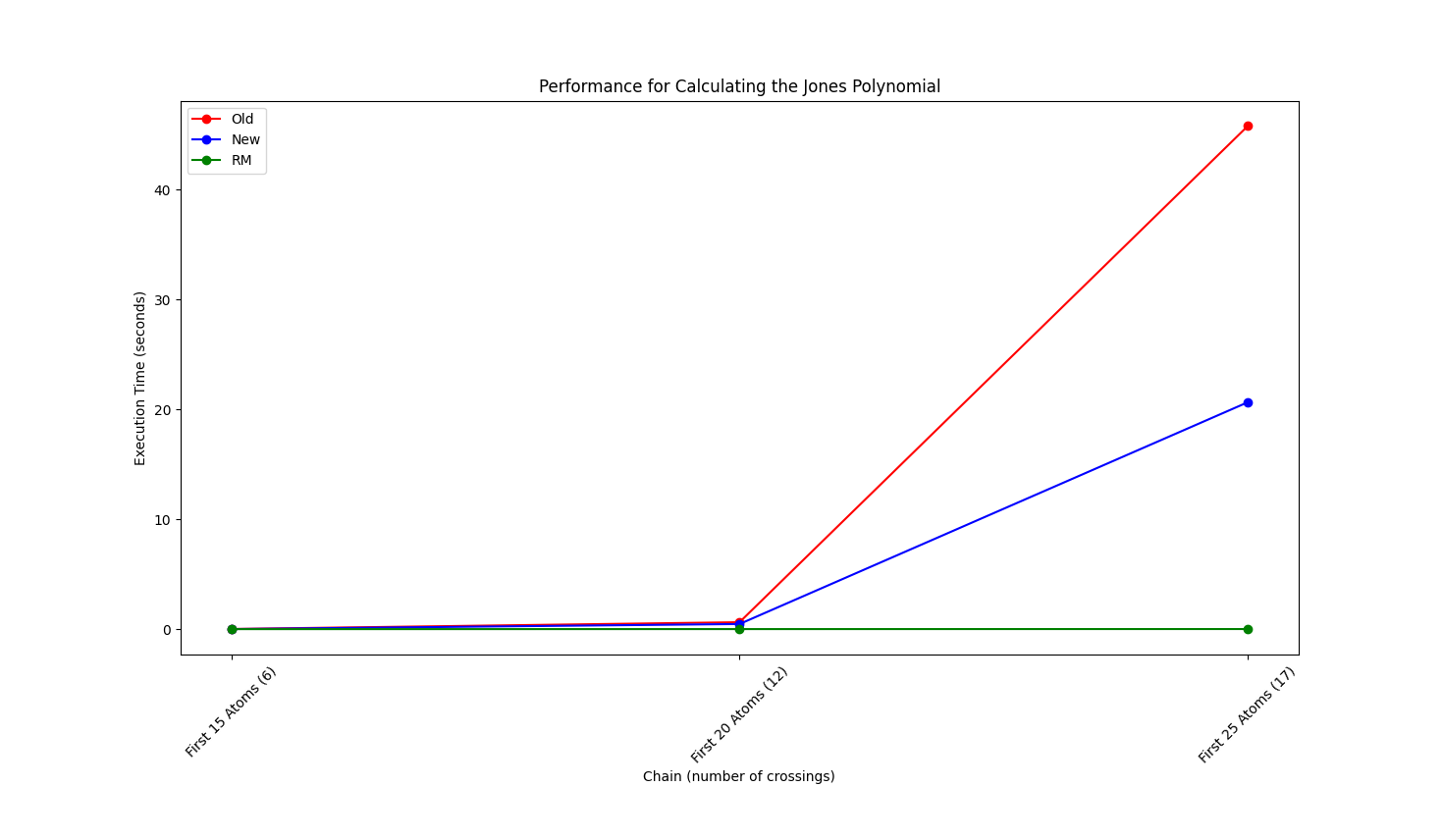}
    \caption{Execution times for structures of increasing length for the different methods.}
    \label{fig:execution}
\end{figure}

\noindent For the first 15 atoms, the old method took 0.01518 seconds, the new method 0.00817 seconds, and the new method with Reidemeister moves 0.00383 seconds. The pattern of decreased execution time with the new methods continues with larger structures. For the first 20 atoms, the old method recorded 0.63376 seconds, the new method showed a substantial decrease to 0.46920 seconds, and the application of Reidemeister moves brought it down to 0.00718 seconds. Finally, for the first 25 atoms, the old method's time was notably higher at 45.79039 seconds, while the new method improved this to 20.65165 seconds, and the inclusion of Reidemeister moves resulted in a dramatic reduction to just 0.01155 seconds. The results seen in \ref{fig:execution} seem consistent with what could have been predicted from our theoretical complexities.

\chapter{Conclusion}
Our computational methods, specifically through the application of divide-and-conquer techniques combined with parallel computing, have notably improved the efficiency of calculating the Jones polynomial. The time cost, once dominated by exponential growth with respect to the number of crossings, has seen a significant reduction. This method combined with recursive Reidemeister moves showed linear runtime activity when tested on the same configurations.
\\ \\
Building upon these successes, we are focusing our efforts on refining the splitting and gluing algorithms for further efficiency. By dividing the problem into even more subdivisions, we aim to reduce the computational workload for each segment, allowing for a more granular and efficient parallelization process. This strategy could potentially lead to a sub-exponential or even polynomial-time complexity in certain cases.
\\ \\
Additionally, the current heuristic for applying Reidemeister moves, while effective, seems far from optimal. There exists the potential to devise a more universally efficient sequence of these moves that could systematically reduce the number of crossings prior to computation. By establishing a more generalized order for applying Reidemeister moves, we could minimize the diagram before the calculation begins, potentially reducing the need for extensive splitting and directly decreasing computational costs.
\\ \\
There is also the need to incorporate these methods back into the original C++ version. The robustness, speed, and portability of C++ make it an ideal candidate for high-performance computing tasks. Our algorithmic improvements in C++ may be better suited to handle large-scale computations more efficiently.
\\ \\

\end{document}